\documentclass[5p,sort&compress]{elsarticle}

\usepackage{hyperref}
\usepackage{epstopdf}
\usepackage{cancel}
\usepackage{subfigure}
\usepackage{amsmath}
\usepackage{amssymb}
\usepackage{latexsym}
\usepackage{tabularx}
\usepackage{graphicx}

\title{Penalized Splines for Smooth Representation of High-dimensional Monte Carlo Datasets}
\author[uw]{Nathan~Whitehorn\corref{cor1}}
\ead{nwhitehorn@icecube.wisc.edu}
\author[uw]{Jakob~van~Santen}
\ead{jvansanten@icecube.wisc.edu}
\author[psu]{Sven~Lafebre}
\ead{s.lafebre@gmail.com}
\address[uw]{University of Wisconsin - Madison, Dept. of Physics,
Madison WI, 53705 USA}
\address[psu]{Pennsylvania State University, Dept. of Physics,
University Park PA, 16802 USA}
\cortext[cor1]{Corresponding author. Tel: +1 608 265 4978}
\journal{Computer Physics Communications}

\begin{document}

\begin{abstract}
Detector response to a high-energy physics process is often estimated by Monte Carlo simulation. For purposes of data analysis, the results of this simulation are typically stored in large multi-dimensional histograms, which can quickly become both too large to easily store and manipulate and numerically problematic due to unfilled bins or interpolation artifacts. We describe here an application of the penalized spline technique \cite{EilersMarxPspline} to efficiently compute B-spline representations of such tables and discuss aspects of the resulting B-spline fits that simplify many common tasks in handling tabulated Monte Carlo data in high-energy physics analysis, in particular their use in maximum-likelihood fitting.
\end{abstract}


\maketitle

\section{Introduction}

Monte Carlo simulation is a common technique for estimating detector response in high-energy physics and tabulated forms of Monte Carlo simulation are commonly used in data analysis as probability density functions in maximum-likelihood fits. As the number of tabulated dimensions increases, numerical problems begin to occur: each bin of the histogram becomes less well filled, statistical fluctuations become larger, zero bins appear. Use of such functions as probability densities with numerical minimization algorithms in maximum likelihood fits can then cause substantial problems with convergence and makes parameter extraction using this approach difficult. The use of penalized spline fits \cite{EilersMarxPspline} in place of raw histograms, however, allows the resolution of most of these problems without resorting to extremely coarse binning and introduces additional capabilities, such as analytic convolution, derivatives, and integrals that can improve fit quality even beyond the removal of numerical instabilities.

The application described here arose in the context of the IceCube detector \cite{daqpaper,amanda_muon_reco}, which uses tabulated Monte Carlo data, among other purposes, for describing photon propagation in glacial ice for simulation of neutrino events, energy reconstruction of muons \cite{photorec}, and for direction, position, and energy reconstruction of electromagnetic and hadronic showers produced by $\nu_e$ and neutral current neutrino interactions \cite{credo}. The heterogeneity of the glacial ice \cite{icepaper} results in a complicated dependence of light propagation on direction, source depth, and receiver depth, resulting in 6-dimensional tables required to describe the light response of the detector to arbitrarily oriented and positioned showers. Similar problems arise in parametrizing detector response for purposes of event reconstruction in many high-energy detectors, either because of variations in the detector medium (as with IceCube) or because of varied instrumentation technology (as in many collider detectors).

A raw table-based approach caused difficulties, especially for directional reconstruction of showers: the tables can require immense amounts of memory (terabytes) and tabulation can introduce numerical artifacts that resulted in seams in the resulting likelihood space and prevented minimizer convergence. These numerical artifacts arise from averaging over a table cell during binning, from inaccurate interpolation between cells, and from statistical fluctuations as the bins are filled. Any attempt to reduce the size of the tables exacerbated the binning artifacts, and vice versa. An interpolating function was the only viable solution to these problems.

An ideal interpolating function for these types of histograms would have several properties: it would be smooth, fast to evaluate, possible to fit deterministically in a small amount of time, extend easily to large numbers of dimensions, and have parameters that are easy to store, while minimizing introduction of bias and artifacts into the fit surfaces. Basis splines (B-splines) fit this description well: a spline of order $n$ has $n-1$ continuous derivatives, the functions can be evaluated quickly, and fitting them to a histogram is a linear problem, so it can be done quickly with standard techniques. The underlying linear problem is inherently sparse, and can be computed efficiently even for very large histograms using Generalized Linear Array Models (GLAM \cite{EilersCurrie}). The linear formulation can also be extended easily to include a regularization term \cite{EilersMarxPspline} that penalizes erratic sweeps in the fit surface in the absence of strong evidence from the data. 

In addition to the general properties listed above, our problem had a number of special requirements that we were able to satisfy through a combination of extensions to the GLAM/penalized-spline fitting technique and optimized routines for the evaluation of tensor-product B-spline surfaces. First, our photon propagation tables were large enough (hundreds of gigabytes) to make it impractical to load the entire 6-dimensional histogram into memory at once, required to perform any fit, so we implemented a method of ``stacking'' spline surfaces fit to 4-dimensional slices of the full histogram to form a full 6-dimensional interpolating spline surface. Second, we needed to assume the same tables for both simulation and reconstruction of events, which have somewhat different required information. In order to simulate events, we needed to be able to sample from the distribution of photon detection times at fixed points in space using its time-differential form, while for maximum-likelihood reconstruction we needed to be able to integrate the distribution of photon detection times over arbitrary intervals as well as efficiently evaluate the gradients of those integrals in all 6 dimensions. We accomplished all three with a single spline surface by fitting the cumulative distribution of photon arrival times, using the technique of non-negative least-squares to enforce monotonicity, and using a basis of B-splines and their derivatives to simultaneously evaluate both the value and gradient of the resulting surface at little additional computational cost. Third, in the maximum-likelihood reconstruction we needed to be able to account for a family of detector effects and event hypothesis uncertainties that can be approximated by a convolution of the photon detection time distribution with Gaussians of various widths. In order to effect such convolutions without repeating the fit once for each kernel, we implemented a method for quickly computing the coefficients of the convolution of the fit surface with an arbitrary B-spline kernel \cite{Stroem19941}.

We have implemented the spline-fitting techniques mentioned above in a library written in C using sparse matrix operations provided by the SuiteSparse \cite{cholmod} and GotoBLAS 2 \cite{gotoblas} libraries, as well as a lightweight library of functions for evaluating the resulting surfaces (code available in the CPC program library). Here we will describe the underlying mathematics of these methods as well as the details of our implementation. 

\section{B-spline surfaces} 
\label{sec:b_spline_surfaces}

To create a smooth approximating function in one dimension, we take a linear combination of B-spline functions defined on a common knot vector $\vec k$, such that the $n$th B-spline of order $m$, $B_m^n$, is defined on the subset of $\vec k$ from $\vec k_n$ to $\vec k_{n+m+1}$.

Functions defined in this way have several attractive properties. Because they are linear combinations of smooth functions, they are likewise smooth, the amount of information required to represent the function is small, and they can be fit to data using standard linear techniques such as Singular Value, QR, or Cholesky decomposition. In addition, because all but $m+1$ of the basis functions are identically zero at any particular point (they have local support), to evaluate the approximation function we need only evaluate the de~Boor algorithm \cite{PracticalGuide} $m+1$ times. This vastly improves computation time for lookups compared to functional representations like Fourier decompositions or polynomial interpolation that require evaluations of basis functions at every point.

\begin{figure}
\centering
\includegraphics[width=\linewidth]{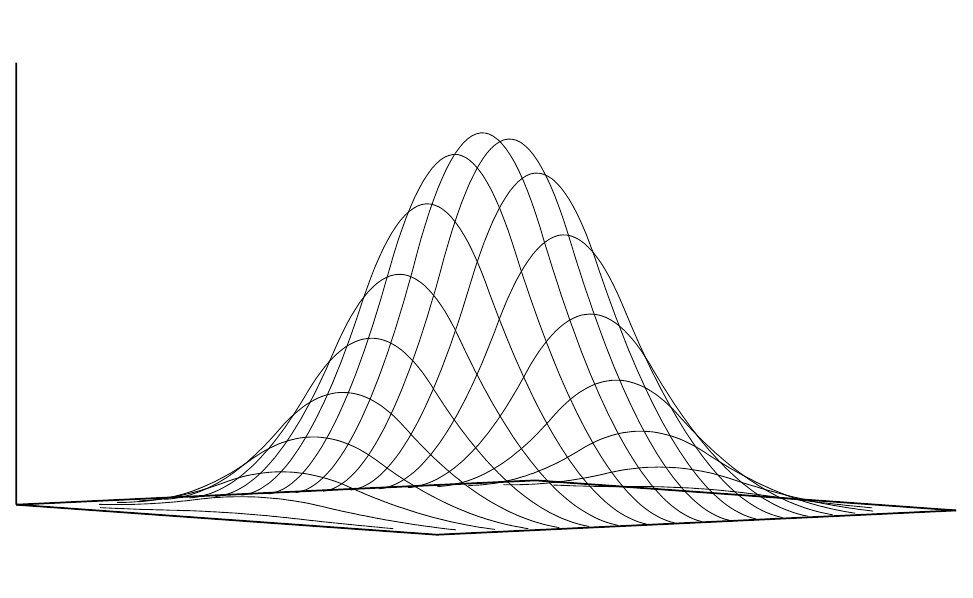}
\caption{A 2-dimensional tensor product spline constructed from two one-dimensional basis splines.}
\label{fig:3dspline}
\end{figure}

We can extend this approach easily into multiple dimensions by adopting the concept of the tensor product spline. Instead of a knot vector, the tensor product basis functions are defined on a rectangular $n$-dimensional knot grid $\overline k$ formed by taking the tensor product of $n$ one-dimensional knot vectors. Each basis function is then a product of one-dimensional B-splines (Figure \ref{fig:3dspline}), which are taken in linear combination as in the one-dimensional case:

\begin{equation}
\label{eq:tensorproduct}
B(\vec x) = \overline \alpha \cdot \overline B = \overline \alpha \cdot \left ( B^i_1(\vec x_1) \otimes B^j_2(\vec x_2) \otimes \dots  \right )
\end{equation}

Because most components of the basis tensor $\overline B$ are zero, as in the one-dimensional case, we recover the same local support property, albeit with more contributing basis functions due to the increased dimensionality of the problem.


\section{Determining B-spline coefficients} 
\label{sec:determining_b_spline_coeffiecients}

\subsection{Ordinary least squares} 
\label{sub:ordinary_least_squares}

The coefficients of the B-spline basis functions can be found by the usual least-squares formulation:

\begin{equation}
\label{eq:normaleq}
B^{\intercal} B x = B^{\intercal} y
\end{equation}

where $x$ is the vector of B-spline coefficients, $y$ the data points in the table, and $B$ a matrix with one column for each basis spline; the rows of $B$ are formed by evaluating the spline at each data point in $y$. The value of $x$ can be found by standard algebraic techniques such as Cholesky decomposition, but the possibility for overfitting and ringing can cause substantial difficulties when using the resulting fits.


\subsection{Penalized Least Squares}

Ringing in the fit surface can be reduced substantially by giving the curve some intrinsic rigidity to follow the general trend instead of following the fluctuations. Using Tikonoff regularization with splines \cite{EilersMarxPspline} allows us to accomplish this. This is implemented in the context of least squares fitting by adding a term to the equations. Instead of solving Equation \ref{eq:normaleq}, we solve the penalized equation:

\begin{equation}
\label{eq:penalizednormal}
\left ( B^{\intercal} B + \lambda P^{\intercal} P \right ) \alpha = B^{\intercal} y
\end{equation}

The solution to this equation now minimizes $|| (B \alpha - y) || + \lambda || P \alpha ||$. $\lambda$ is then the strength of the regularization (called the smoothing parameter hereafter). For $\lambda = 0$ we recover the unpenalized normal equations \eqref{eq:normaleq}. For $\lambda \to \infty$, the regularization term dominates, and we end up with a curve of the shape specified by the regularization term, which is a form of Bayesian prior (for order-2 Tikonoff regularization, this will be a straight line). Details of the form of this matrix can be found in \cite{EilersMarxPspline}.

\subsection{Penalized Least Squares in Multiple Dimensions}
\label{sec:glam}

In principle, the extension of least squares fitting to multiple dimensions is easy. As before, we have a set of ($n$-dimensional) basis functions $B$, to be fit at a set of $n$-dimensional positions to a measured set of data $y$. With some smoothness constraint on each dimension, we then want to minimize

\begin{equation}
||B \alpha - y|| + \lambda_1 || P_1 \alpha || + \lambda_2 || P_2 \alpha || + \cdots + \lambda_n || P_n \alpha ||
\end{equation}

In analogy to Equation \ref{eq:penalizednormal}, we can write down the corresponding system of linear equations:

\begin{equation}
\label{eq:multidimnormal}
\left (B^{\intercal}B + \sum_i^n \lambda_i P_i^{\intercal} P_i \right ) \alpha = B^{\intercal} y
\end{equation}

The difficulty here arises that $B$ can be infeasibly large in the multi-dimensional case. For 25 knots on each of 4 axes, and a table with 200 million cells (typical numbers for a single IceCube photon propagation table), $B$ will have dimension 390625 by 200 million and require 568 TB of RAM.

There are two important things to notice: the first is that $B$ is quite sparse. Because of the local support property of B-splines, most of the basis functions have no support at most of the data points. As such, we can eliminate all of the zeroes from the matrix using sparse-matrix routines. This reduces memory consumption in the above problem by a factor of 100.

The second thing is that $B$ is not present in isolation in Equation \ref{eq:multidimnormal}, but instead only occurs as the combination $B^{\intercal}B$. Whereas $B$ has dimensions of the number of basis functions by the number of data points, $B^{\intercal}B$ is a square matrix of side length the number of basis functions. For the example problem above, being able to work directly with $B^{\intercal}B$ without the need to ever explicitly form $B$ would reduce memory consumption by another factor of 500. This also means, critically, that neither the memory requirements nor the CPU requirements of this fitting algorithm would depend on the number of data points, but instead only on the number of basis functions. Working directly with $B^{\intercal}B$ can be accomplished using GLAM \cite{EilersCurrie}.

\subsection{Table Stacking and Pseudointerpolation}

For each run of the IceCube photon simulation \cite{Lundberg:2007p2} we intend to fit, we produce 4-dimensional light distributions for a single source configuration. For a table-based simulation, many source configurations are simulated and the distributions for an arbitrary source produced from interpolation between the tables for similar sources. For spline tables, we wish to be able to do this as well. The easiest solution, to do a global fit, is impracticable even using GLAM \cite{EilersCurrie}. In addition, because there is no bin averaging between sources, smoothing (and thus least-squares fitting) is unnecessary, and we can stick to interpolation algorithms.

What we do then is to ``stack'' the individual 4-dimensional tensor-product spline surfaces into a single 6-dimensional tensor-product spline surface. The normalization of the spline basis functions is such that we can simply stack the fit coefficients of each sub-table while creating a knot vector such that the original values of the sub-tables form control points for the interpolating splines in the extra dimensions.

Given a knot vector $\vec k$, the maxima of the B-splines defined on that vector can be found by solving the equation for the first derivatives of each spline \cite[Chapter~X,~Equation~12]{PracticalGuide}. Given knots spaced at equal intervals $\Delta \vec k$, these also are the centers of each basis function and can be found at:

\begin{equation}
\label{eq:splinecenters}
\vec B_n^{max} = \vec k - \frac{n - 1}{2} \Delta \vec k 
\end{equation}

By inverting Equation \ref{eq:splinecenters}, we can acquire a formula for a knot vector given a vector of centers $\vec x$:

\begin{equation}
\label{eq:stackingknots}
\vec k = \vec x + \frac{n - 1}{2}  \Delta \vec x
\end{equation}

We must also add points to the knot vector, since a spline function of order $n$ made of $m$ basis functions (and so $m$ coefficients) must be defined by $m+n+1$ knots. Thus we need $n+1$ extra knots, of which the first $n$ are placed at linearly extrapolated positions before the positions from Equation \ref{eq:stackingknots}, and one placed at a linearly extrapolated position after the last element.

\begin{figure}
\centering
\includegraphics[width=\linewidth]{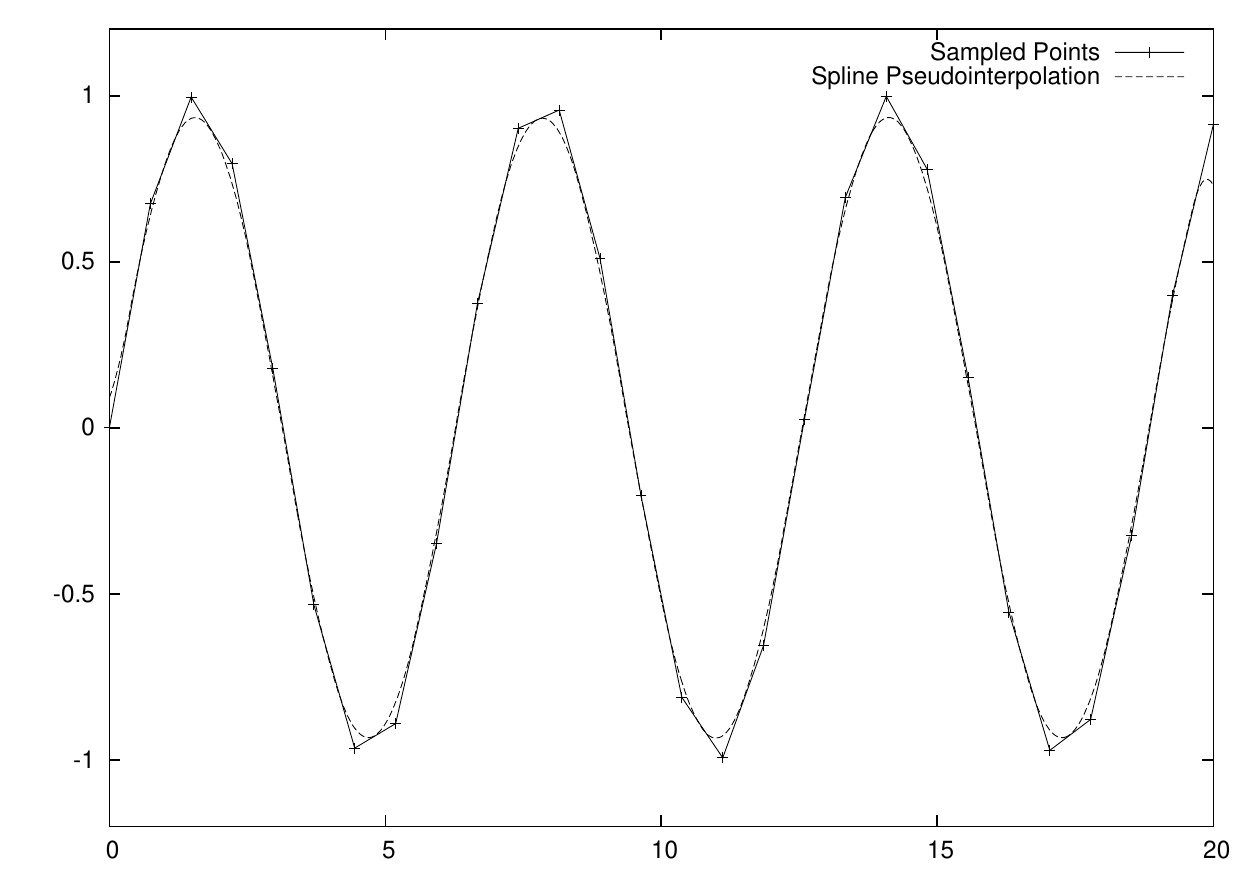}
\caption{Pseudointerpolation of a set of points sampled from $\sin(x)$ using the same algorithm as for table stacking. Since B-splines do not interpolate their control points, but instead maximally fill the convex hull formed by those points, the interpolation algorithm will fall slightly short of extrema, while generally maintaining accuracy.}
\label{fig:convexhull}
\end{figure}

Iterative application of Equation \ref{eq:stackingknots} with stacking of coefficients will then produce a new tensor product spline surface with the sub-table locations as control points. It is worth noting that this surface does not exactly interpolate the original tables, as B-splines do not interpolate their control points. Instead, it forms a curve that maximally fills the convex hull of the set of control points (Figure \ref{fig:convexhull}). In general, this is an excellent approximation to interpolation, but sharp extrema will be systematically underestimated by a small amount.

\section{Cumulative B-spline surfaces} 
\label{sec:cumulative_b_spline_surfaces}

Using a single B-spline surface to represent the distribution of photon arrival times
for both simulation and reconstruction purposes posed a special challenge. To be able
to efficiently evaluate the integral of the distribution over arbitrary time windows,
we fit the cumulative distribution to a tensor-product B-spline surface, using a
modified version of the penalized-spline formulation \cite{EilersMarxPspline} to
enforce the monotonicity of the surface. In order to recover the differential
distribution and other elements of the gradient from the cumulative form, we
implemented an efficient method of simultaneously evaluating the value and gradient
of a tensor-product spline surface. Lastly, we implemented a method of quickly -- and analytically --
convolving a spline surface with a spline kernel in order to approximate detector
effects such as PMT transit-time spread during reconstruction.

\subsection{Enforcing monotonicity} 
\label{sub:enforcing_monotonicity}

\begin{figure}[htb]
	\centering
		\includegraphics[width=\linewidth]{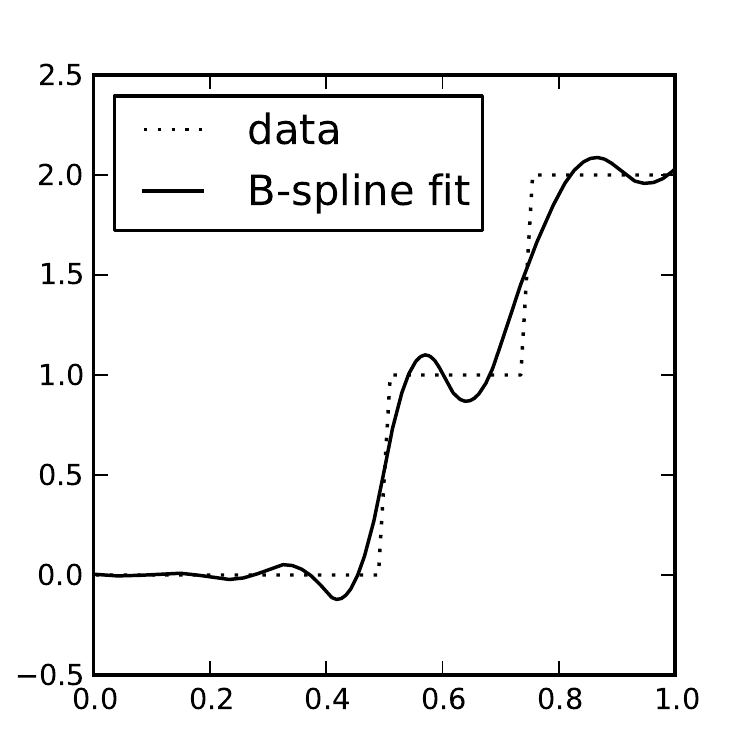}
	\caption{An example of the ``ringing'' that can occur as an artifact of least-squares fitting with B-splines. The sharp transitions are unrepresentable in the B-spline basis, so the spline surface must fluctuate around the data in order to minimize the residual. This can cause the spline surface to become negative where the data are strictly positive and non-monotonic where the data are strictly monotonic.}
	\label{fig:illustration:ringing}
\end{figure}

While the B-spline surface determined by the least-squares condition is by
definition the closest fit to the data, it does not necessarily share some
desirable properties with the data. In particular, when the data can't be exactly
represented as a spline surface, the fit surface may ring (Figure~\ref{fig:illustration:ringing}). This
can cause the fit surface to become negative in a region where the data are
strictly positive or non-monotonic where the data are strictly monotonic. Although this behavior can be substantially reduced by regularization, it is critical that at least monotonicity be exactly obeyed for our application, where we fit for the cumulative
photon arrival time distribution, which must by definition be everywhere monotonic.
It is possible, however, to express the monotonicity constraint along a single
dimension (time, in our case) without substantial modification to the usual least-squares fitting procedure.

We implemented the method suggested by \cite{beliakov2000shape}, in which we
transform one axis of the tensor-product B-spline basis to a cumulative
T-spline (``Trapezoidal'' spline)
basis in which each basis function is the sum of the B-spline basis functions following it.
The equation for the coefficients of the penalized B-spline surface \eqref{eq:multidimnormal}
can be converted to the T-spline basis by multiplying the basis and penalty matrices from
the right with an upper-triangular matrix:

\begin{equation}
	\tilde{B} = BU = B
	\begin{pmatrix}
	1      & 1      & \cdots & 1      \\
	0      & 1      & \cdots & 1      \\
	\vdots & \vdots & \ddots & \vdots \\
	0      & 0      & \cdots & 1      \\
	\end{pmatrix}
\end{equation}

Unlike the B-splines from
which they are derived, the T-spline basis function have highly non-local
support; e.g. the first T-spline basis function has support over the entire knot
field. This makes evaluating the spline surface arduous; in the extreme case of a
point in the support of the last T-spline basis function, evaluating the spline
surface involves all the preceding coefficients. Luckily, we can use the T-spline
basis for fitting only, transforming the T-spline coefficients into the
corresponding B-spline coefficients before storing them.

The T-spline basis, however, makes it particularly easy to enforce monotonicity
in one dimension. In this basis, the spline surface is monotonic if and only if all the
T-spline coefficients are positive. While not quite as simple as the
unconstrained problem, there exist algorithms to solve the non-negative least-squares problem as a series of unconstrained problems in a finite number of steps.

Given $A \in \mathbb{R}^{M \times N}$ and $b \in \mathbb{R}^{N}$, $x \in
\mathbb{R}^{M}$ is the solution to the non-negative least squares problem:

\begin{equation}
	\min_{x} || Ax - b ||^2_2 : x \geq 0
\end{equation}

The Karush-Kuhn-Tucker optimality conditions lead to the following linear
complementarity problem \cite{Portugal:1994}:

\begin{equation}
	\label{eq:LCP}
	y = A^{T}Ax - A^{T}b : y \geq 0, \,\, x \geq 0, \,\, x^{T}y = 0
\end{equation}

An active set approach is commonly used to find the solution to
\eqref{eq:LCP}. The set of coefficients $x$ is partitioned into a free set
$\mathcal{F}$ and a constrained set $\mathcal{G}$ (often called the passive and active sets, respectively).
The coefficients in the free set
are the solution to the unconstrained least squares problem in the subspace
of the free set, while those in the constrained set are clamped to zero. The
problem then reduces to quickly and efficiently finding the partition of the
coefficients that satisfies \eqref{eq:LCP}. The solution can be found
iteratively, but unlike the minimization of a general function, it is possible to
construct algorithms that are provably finite.

\begin{figure}[htb]
	\centering
	\includegraphics[width=\linewidth]{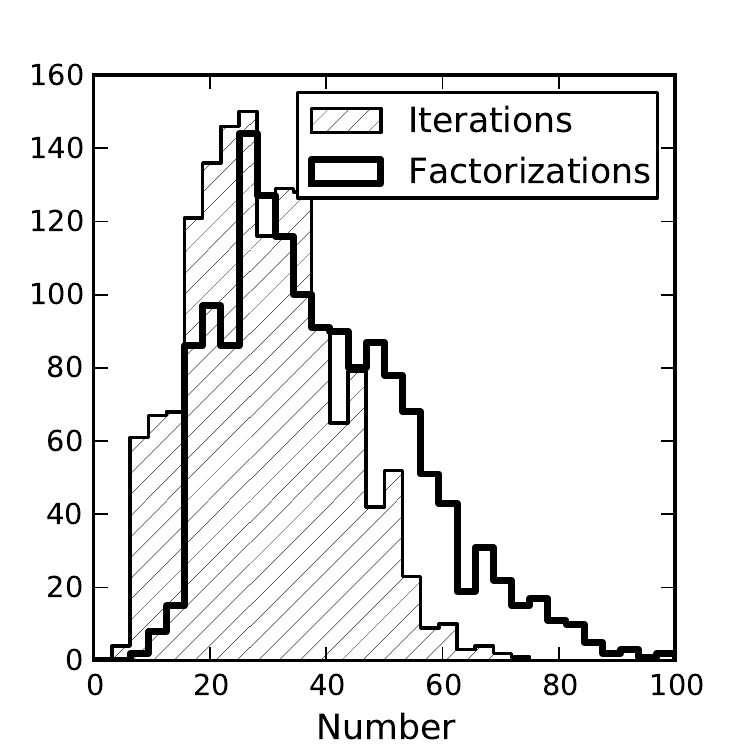}	
	\caption{Number of iterations and complete Cholesky factorizations needed for the BLOCK3 algorithm to converge when fitting large multi-dimensional photon tables (100 million table cells and 400 thousand spline coefficient).}
	\label{fig:illustration:performance:block3_iterations}
\end{figure}

\begin{figure}[htp]
	\centering
	\subfigure[Computation time in CPU hours]{
	    \label{fig:illustration:block3_performance:cpu_time}
	    \includegraphics[width=0.4\linewidth]{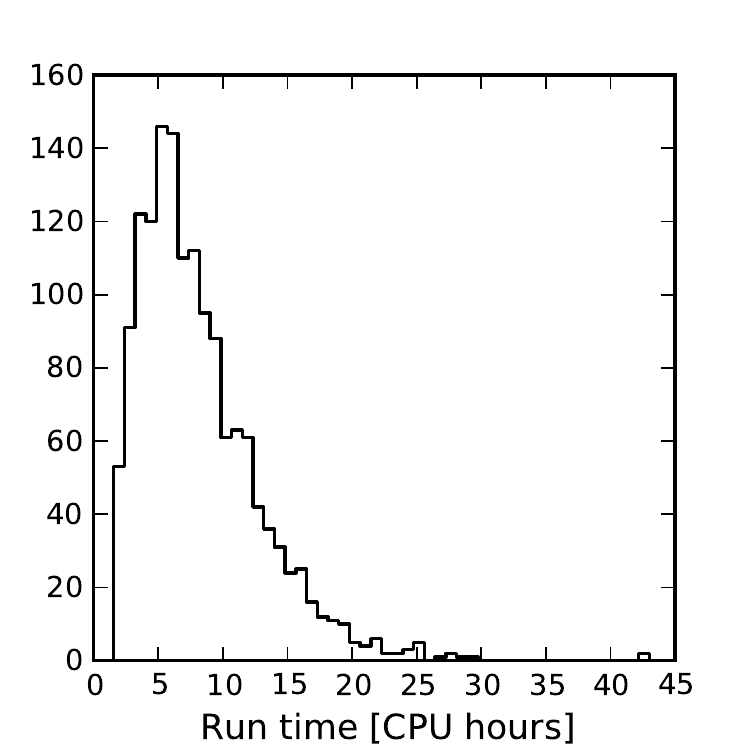}}
	\subfigure[Computation time in real hours]{
	    \label{fig:illustration:block3_performance:wallclock_time}
	    \includegraphics[width=0.4\linewidth]{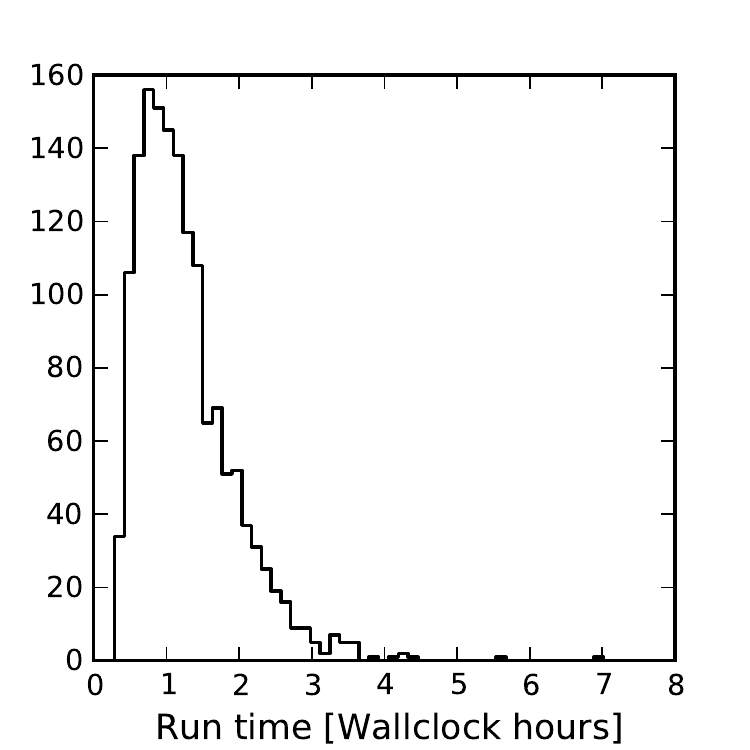}}
	\subfigure[Average CPU utilization]{
	    \label{fig:illustration:block3_performance:percent_cpu}
	    \includegraphics[width=0.4\linewidth]{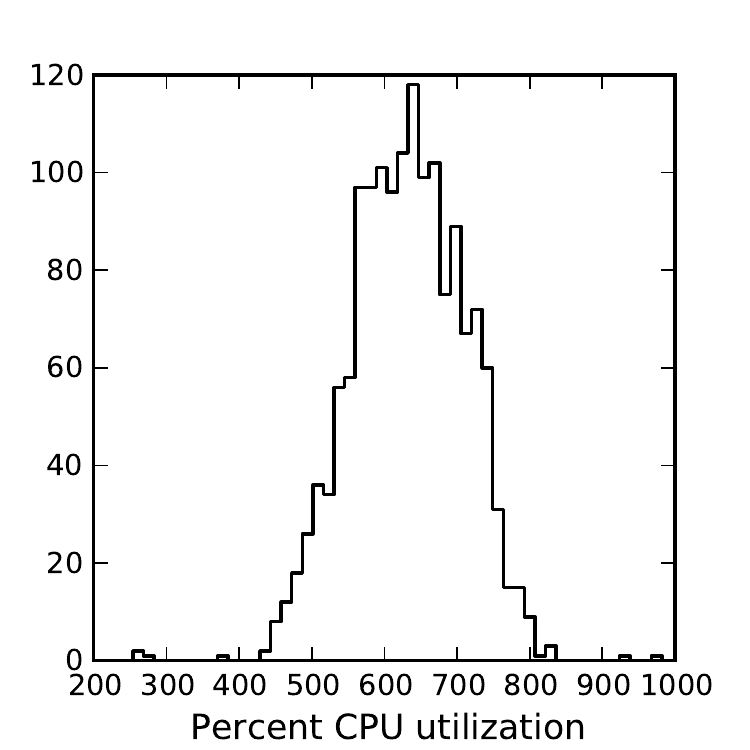}}
	\subfigure[Memory usage sampled every 30 seconds over 400 hours of fitting. ]{
	    \label{fig:illustration:block3_performance:mem}
	    \includegraphics[width=0.4\linewidth]{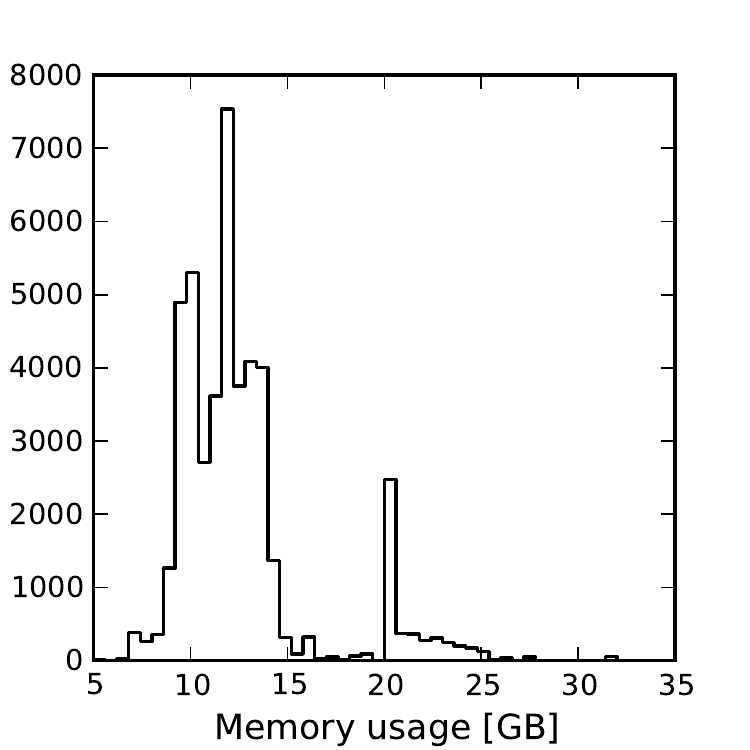}}
	\caption{Resource consumption of penalized spline fits. Shown are two extremely large fitting jobs (100 million table cells and 400 thousand coefficients) running in parallel on a
	    16-core, 2.3 GHz AMD Opteron system with 64 GB of memory using SuiteSparse \cite{cholmod} and GotoBLAS 2 \cite{gotoblas}. The memory usage peak at 20 GB represents the memory needed for the initial factorization of the full $A^TA$, while the broader, lower-memory peak represents the memory needed for the factorization of only the unconstrained parts of $A^T A$.}
	\label{fig:illustration:block3_performance}
	\end{figure}

The best known of these is the Lawson-Hanson NNLS algorithm \cite{Lawson:1974}. For very large problems, NNLS can be slow to converge, so we additionally implemented and examined two block-pivoting algorithms, the Portugal/Judice/Vincente (PJV) algorithm \cite{Portugal:1994} and Adlers BLOCK3 \cite{Adlers:1998}, all using sparse matrix operations. Both provide substantial speed improvement relative to NNLS, but, due to occasional cycling behavior in the PJV algorithm, we recommend use of BLOCK3 for all applications. BLOCK3 provides rapid convergence even for very large problems (Figure \ref{fig:illustration:performance:block3_iterations}, \ref{fig:illustration:block3_performance}) and converges deterministically in all circumstances.

\subsection{Gradients of spline surfaces} 
\label{sub:gradients_of_spline_surfaces}

Even though we fit the cumulative distribution of photon arrival times, we still need the
probability density functions for sampling as well as the other dimensions of the gradient for accelerating
minimizer convergence in maximum-likelihood fitting. Using the de~Boor recursion relation 
\cite[Chapter~X]{PracticalGuide} and the multiplication rule, we can easily construct a
new set of basis functions that are the derivatives of B-splines:

\begin{align}
B_n^{i '}(x) = & \frac{(x - \vec k_i) B_{n-1}^{i'}(x) + B_{n-1}^i(x)}{\vec k_{i+n} - \vec k_i} + \\
   & \frac{(\vec k_{i+n+1} - x)B_{n-1}^{i+1 '}(x) - B_{n-1}^{i+1}(x)}{\vec k_{i+n+1} - \vec k_{i+1}}
\end{align}

We can unpack one of these recursions in order to write the derivatives as simple
linear combinations of two lower-order B-splines:

\begin{equation}
	\label{eq:debooralgo_deriv}
	B_n^{i '}(x) = \frac{n}{\vec k_{i+n} - \vec k_{i} }B_{n-1}^{i}(x) -  \frac{n}{\vec k_{i+n+1} - \vec k_{i+1} }B_{n-1}^{i+1}(x)
\end{equation}
The lower-order basis functions needed to form the derivative basis can be obtained at
almost no additional computational cost by pausing the bottom-up calculation of $B_n^i$ 
\cite[Algorithm BSPLVB]{PracticalGuide} at order $n-1$ before proceeding to the final step.

Because differentiation and addition commute, we can interchange these
differential spline basis functions on the axis of interest with the original
ones, while keeping all the other axes and the coefficient table unchanged.
Keeping the coefficients unchanged also allows us to efficiently evaluate the
spline surface and its gradient using SIMD operations. In our case, evaluating
the spline surface and the 6 elements of its gradient in sequence takes 7 times
as long as a single evaluation, while the same computation implemented in terms
of operations on 4-element basis vectors takes slightly less than twice as long.


\subsection{Analytic convolution of spline surfaces} 
\label{sub:analytic_convolution_of_spline_surfaces}

While scattering processes account for the vast majority of the time-delay
distribution in detected photons, it can be important to account for detector
effects as well when reconstructing events. In particular, the propagation of
current pulses through the photomultiplier tube and inter-DOM clock synchronization \cite{daqpaper}
introduce small timing uncertainties in the measured photon arrival time distribution.
Also, while electromagnetic and hadronic showers
are approximately point-like on the scale of the inter-string spacing in IceCube,
they do have some spatial extent. Both of these effects can be approximately
accounted for by convolving the photon delay-time distribution with a Gaussian.

\begin{figure}[htb]
	\centering
		\includegraphics[width=\linewidth]{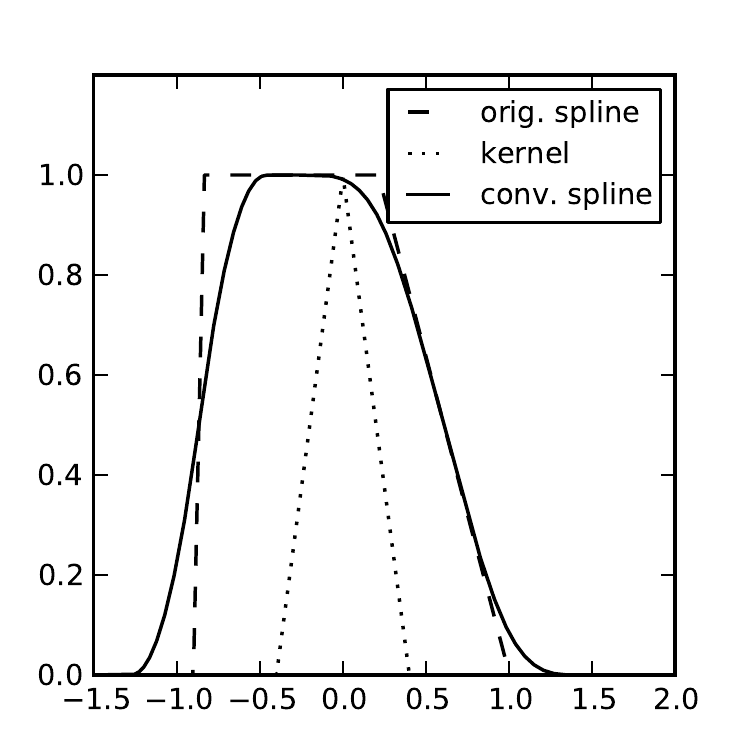}
	\caption{An example of spline convolution. Here, an order-1 spline surface on a
	    non-uniform knot field is convolved with an order-1 spline, resulting in an
	    order-3 spline surface.}
	\label{fig:illustration:spline_convolution}
\end{figure}

On a uniform knot field, a B-spline basis function of order $n$ is proportional
to the $n$-fold self-convolution of a constant function between two adjacent
knots, and so the $n$th order uniform B-spline becomes
proportional to a normal distribution as $n \to \infty$. B-splines of relatively
low order can be used to approximate Gaussian convolution kernels, albeit with
limited support. Furthermore, it is possible to analytically convolve a spline
surface with a B-spline kernel and expand the result in a new B-spline basis (Fig. \ref{fig:illustration:spline_convolution}).
These are useful properties, as they allow convolution of a spline surface by a  pseudo-Gaussian kernel merely by manipulating its coefficients.

The procedure for analytically convolving splines is described in
\cite{Stroem19941} in a slightly different form than we require for our
application, and we will review our adaptation of it here for the sake
of clarity. As before, we will use $B_m^i$ to denote the $i$th B-spline of
order $m$ on a knot vector $\vec{t}$ with the usual de~Boor normalization
such that the set of splines on $\vec{t}$ sum to 1 everywhere. The convolution
kernel will be represented by a spline $M_n^j$ on a knot vector $\vec{\tau}$, but
normalized so that each individual basis function integrates to 1.

The convolution of $B_m^i$ and $M_n^j$ can be written\footnote{This is a
variation on Equation~13 of \cite{Stroem19941}.} in terms of dummy variables
$x$ and $y$ as

\begin{align}
	(B_m^i \star M_n^j)(z) & = (t_{i+m+1} - t_{i}) \frac{n! (m+1)!}{(m+n+1)!}
	    \notag \\ 
	    & \times [t_i, \dots, t_{i+m+1}]_x [\tau_j, \dots, \tau_{j+n+1}]_y \notag \\
	    & \times (x + y - z)_{+}^{m+n+1}
\end{align}

where $[t_i, \dots, t_{i+k}]_x$ is the divided-difference operator with
respect to $x$ and $(\cdot - z)_{+}^n$ is the truncated power function

\begin{equation}
	(a - b)_{+}^n = 
	\begin{cases}
		(a-b)^n & b < a \\
		0 & b \geq a
	\end{cases}
\end{equation}

Furthermore, $B_m^i \star M_n^j$ is a B-spline surface of order $m+n+1$ on the
combined knot field
$\vec{\rho} \equiv \{t_i,\dots,t_{i+m+1}\}\uplus\{\tau_j,\dots,\tau_{j+n+1}\}$\footnote{The combined
knot field is formed by $m+2$ copies of $\{\tau_j,\dots,\tau_{j+n+1}\}$, each
centered on an element of $\{t_i,\dots,t_{i+m+1}\}$. For example,
$\{0,1,2\}\uplus\{-0.1, 0, 0.1\} = \{-0.1, 0, 0.1, 0.9, 1, 1.1, 1.9, 2, 2.1\}$.
For a proof that $B_m^i \star M_n^j$ is contained in the spline space of
order $m+n+1$, see Theorem 8 of \cite{Stroem19941}.}. Given this fact and the
ability to evaluate $(B_m^i \star M_n^j)(z)$ for arbitrary $z$, we can expand
the convolution in its B-spline coefficients.

A B-spline surface of order $m$ defined on a knot field $\vec{\rho}$ is a linear
combination of B-splines:

\begin{equation}
	f(x) = \sum_i a_i B^i_{m,\vec{\rho}}
\end{equation}

The coefficients of the B-spline expansion are given by

\begin{equation}
	\label{eq:blossom_coefficients}
	a_i = \mathcal{B}(f(\xi_i))(\rho_{i+1},\dots,\rho_{i+m})
\end{equation}

where $\xi_i \in [\rho_i,\rho_{i+m})$ and $\mathcal{B}$ is the
multilinear blossom of $f$ \cite{Ramshaw:1987}. The multilinear blossom is
a function that is linear in each of its $m$ arguments and is identical to $f$
when evaluated on its diagonal, that is, when all its arguments are identical.
The blossom of $B_m^i \star M_n^j$ is given by

\begin{align}
	\label{eq:convoluted_blossom}
	\mathcal{B}((B_m^i \star M_n^j)(z))(\rho_{i+1},\dots,\rho_{i+m+n+1}) =
	    (t_{i+m+1} - t_{i}) \notag\\
	    \times \frac{n! (m+1)!}{(m+n+1)!}
	    [t_i, \dots, t_{i+m+1}]_x [\tau_j, \dots, \tau_{j+n+1}]_y \notag\\
	    \times \Theta(x + y - z) \prod_{l=i+1}^{i+m+n+1}(x + y - \rho_l)
\end{align}

where $\Theta$ is the Heaviside step function.

In our application the convolution kernel is given by a single unit-normalized spline $M_n^1$,
and we can apply \eqref{eq:blossom_coefficients} to write the coefficients $\vec{a}$ of
the B-spline expansion of the convolved surface in terms of the coefficients $\vec{b}$
of the un-convolved surface as

\begin{equation}
	\vec{a} = T \vec{b}
\end{equation}

where $T$ is a matrix whose elements can be found by evaluating the blossom
defined in \eqref{eq:convoluted_blossom}:

\begin{equation}
	T_{ij} = \mathcal{B}((B_m^j \star M_n^1)(\rho_i))(\rho_{i+1},\dots,\rho_{i+m+n+1})
\end{equation}

This is particularly useful for tensor-product spline surfaces: the
convolution can be carried out along a given dimension by calculating the
transformation matrix once and then applying it to each slice of the coefficient
grid along that dimension.

\section{Conclusion} 
\label{sec:conclusion}

Penalized splines provide a number of attractive properties for handling detector-response simulations in high-energy physics experiments where alternative parametrizations are not practicable: fast and efficient deterministic fitting, even for very large datasets, integrated smoothing and extrapolation, and the ability to easily perform a variety of mathematical operations. The conventional penalized spline technique \cite{EilersMarxPspline} was extended here with new tools and with existing algorithms \cite{Stroem19941, Lawson:1974, Adlers:1998} for convolution of the resulting spline tables and use with datasets typical in high-energy physics.
Use of multi-dimensional tabulated Monte Carlo data in maximum-likelihood fits can be critical for many complicated situations such as detailed particle interaction reconstruction in non-segmented inhomogeneous detectors such as IceCube \cite{credo}.
Most other tabulation strategies (multilinear interpolation, polynomials) quickly become intractable as the dimensionality of the problem increases due to numerical issues or computational footprint, rendering problems that can be easily solved in principle impossible or extremely difficult in practice.
Use of spline surfaces, along with the ability to perform on-the-fly differentiation, integration, and convolution, solves many of these problems and greatly improves the ability to use such tables in maximum likelihood fits.
Similar problems of detector response parametrization occur across high-energy physics; use of penalized splines may help with related problems in many detectors.


\section*{Acknowledgments}

We would like to thank Brian Marx and Paul Eilers for invaluable assistance with implementation of the fitting algorithms. J. van Santen was supported under NSF grant 0636875 and N. Whitehorn by the NSF GRFP. Computing resources provided with support from the NSF Office of Polar Programs and Physics Division as well as from the U. of Wisconsin Alumni Research Foundation. Thanks to S. BenZvi for packaging the fitting software for external use.

\bibliographystyle{elsarticle-num}
\end{document}